\documentclass{article}
\usepackage[active]{srcltx}

\newtheorem{theorem}{Theorem}
\newtheorem{lemma}[theorem]{Lemma}
\newtheorem{corollary}[theorem]{Corollary}
\newenvironment{proof}{{\it Proof.\ \/}}{\  }
\newcommand{\la}{\leftarrow}

\newcommand{\ra}{\rightarrow}
\newcommand{\TM}{{\sc TM}{} }
\newcommand{\NTM}{{\sc NTM}{} }

\newcommand{\LINE}{\noindent\hrulefill}

\begin{document}

\thispagestyle{empty}
\pagestyle{plain}
\title
{Is Space a Stronger Resource than Time?\\
Positive Answer for the Nondeterministic \\At-Least-Quadratic Time Case }

\author{
    Nicola Caporaso\\
    \footnotesize    CATTID, Universit\`a ``Sapienza" di Roma\\
    \footnotesize    Piazzale Aldo Moro 5, 00185 Roma, Italy\\
    \footnotesize    Email: \texttt{nicola.caporaso@uniroma1.it}
}

\maketitle

\begin{abstract}
We show that all languages accepted in time $f(n)\geq n^2$
can be accepted in space $O(f(n)^{1/2})$ \textit{and} in time $O(f(n))$.
The proof is carried out by simulation, based on the idea of
\textit{guessing} the  sequences of internal states of the simulated TM
when entering  certain critical cells, whose location is also guessed.
Our method cannot be generalised easily to many-tapes TMs.
And in no case can it be relativised.
\end{abstract}

\section{Introduction}


Let $T_M(n)$ and $S_M(n)$ denote the time and space consumed by a Turing Machine (\TM)
$M$ which, given an input of length $n$, stops operating.
Now, assume that $M$ is an acceptor for the language  $L=L(M)$.
From the \textit{linear} space-compression theorem, for all constants $c$,
one can find a new \TM $M^*$ such that $L=L(M^*)=L(M)$
and
\begin{equation}
    cS_{M^*}(n)\leq T_{M^*}(n)=T_M(n).
\end{equation}
One might ask whether a better than linear result can be obtained.
This is not a trivial question: after all, \textsc{p$\stackrel?=$pspace} is a major problem in computer
science. The nondeterministic case is equally interesting, given that
\textsc{np$\stackrel?=$npspace} is a major problem too.


We will prove the following
\begin{theorem}\label{theorem}
    For every \NTM $M$, another \NTM $M^*$ and a constant $a$
    can be defined such that, for all input $w$ and $n\geq|w|$,
    $M^*$ accepts $w$ in time $n^2$ and space $n$
    if and only if $M$ accepts $w$ in time $an^2$.
\end{theorem}
This allows us to answer positively the question
in the case of single-tape nonderministic \TM (\NTM)
at and above the quadratic time level.
The following
\begin{corollary}
    \[
        \mbox{\textsc{one-tape-ntime}}(f(n))=
        \mbox{\textsc{one-tape-ntimespace}}(f(n);f(n)^{1/2})
    \]
\end{corollary}
is the main result of this paper.

We don't see an easy way to extend the result to many-tapes TMs.
We would like to stress that this result cannot be extended to oracle-TMs either,
because one cannot put an upper bound on queries to the oracle.
One might speculate on the interest of such  non-relativisable arguments
in investigations on the separation problems.

The evaluation of the price (in terms of time) to be paid to save space
is a topic of complexity theory that was initiated by Hopcroft and Ullman,
who proved that deterministic and nondeterministic single-tape TMs
respecting a time bound $T(n)$ can be simulated in space $T^{1/2}$ within
a time exponential in $T(n)$ \cite{HopUll68}.
Ibarra and Moran \cite{IbarraMoran}
proved that single-tape  TMs whose runtime is bounded above
by $T(n)$ can be simulated in  time $T(n)^{3/2}$ and space $T(n)^{1/2}$.
As far as we know, however, \textit{free-of-charge } results have not been proved so far.
We show that not too long crossing sequences exist by a method
that we have derived from from \cite{HopUll68}.

\section{Definitions}

We will introduce a \NTM $M$ with a single half-infinite tape.
The tape is partitioned in blocks, all except at most one of
the same length $n$.
The \NTM will visit each block a certain
number of times: we call each of these visits a phase.
The sequence of all the visits $M$ makes on a given
block is called the block's history.
In the following section, we will see how $M^*$ works by
trying to guess a possible story for the operation of $M$
until it arrives at the correct one.

Let us fix, for the remaining part of this paper,  a \NTM $M$,
an input $w$ for $M$, and a number $n\geq|w|$.
Let us identify the states of $M$ with the numbers  $0,1,\ldots$
Some states are deterministic, while others are not.
Without any loss of generality (see for example \cite{HU79}, chapter 7)
we may assume that
\begin{enumerate}
    \item The tape is infinite to the right. We call \textit{cell} $h$
the $h$-th cell ($h\geq 1$), counting from the left end of the tape.
We use $\Delta$, often with affixes, as a variable defined on
$\{-1,+1\}$. This variable will be used to identify the direction
of motion of $M$ by understanding $-1$ to mean \textit{left},
and $+1$ to mean \textit{right}.

    \item When in a deterministic state, $M$  either moves in the direction $\Delta$,
or else it writes on its tape, \textit{but not both}.
If it tries to move left from  cell 1, then it stops operating
(but it may stop in other ways too).
When in a nondeterministic state, it
just chooses among a number $>1$
of next states, \textit{but it does not move  or write}.

    \item $M$ starts operating in the \textit{initial} state $0$, with
$w$ stored in the cells $1,\ldots,|w|$.
To accept, it tries to move left from  cell 1 in the (only) \textit{accepting} state  $1$.
\end{enumerate}

We have a \textit{computation} $C$
for each sequence of nondeterministic choices made by $M$ on $w$.
The time for $C$ is the number of moves it includes, and its
space is the number of distinct cells it visits.
$M$ accepts its input within time $h$ and space $k$ if
there is a computation that takes time $h$ and space $k$.
Other computations may accept the input in time $h^*>h$
and/or space $k^*>k$, reject it, or never halt.

We will call $\beta_i$ the \textit{boundary}  between cells $i$ and $i+1$.
We will focus on the behaviour of $M$ at evenly spaced boundaries,
starting at $\beta_P$, with spacing $n$. Accordingly,
for each $P\leq n$ we define a \textit{partition}
$\pi_P$ of the first $n^2$ cells into \textit{blocks}
in the following way: the block $B_1$
consists of the first $P$ cells, and $B_{j>1}$
consists of the $n$ cells from  $P+(j-1)n+1$ to $P+jn$.
We will call the boundary between two adjacent blocks $B_{j}$ and $B_{j+1}$,
a \textit{milestone} $\mu_j$; clearly, $\mu_j = \beta_{P+jn}$.
In addition, we will call $\mu_0$ the left end of the tape.

For a given computation,
let a \textit{phase} denote the behaviour of $M$ during a single visit to a block,
until it either stops operating without leaving the block, or it moves across a milestone.
Phase 1 goes from the start to when $M$ leaves for the first time the block $B_1$
to enter, from the left, $B_2$.
If by the end of phase $k$,  $M$
leaves $B_j$ moving in the direction $\Delta$, then phase $k+1$
is the period of operation of $M$ on $B_{j+\Delta}$ until $M$
leaves it to come back to $B_j$, or to enter $B_{j+2\Delta}$.

A \textit{descriptor} is a 4-ple $D=(p,j,i,\Delta)$ saying that, at the beginning of phase $p$,
$M$ is in state $i$, and moves across $\mu_j$ in the direction $\Delta$.
We adopt the following conventions:
\begin{enumerate}
    \item  We will sort descriptors by phase number into \textit{sequences}.
    A sequence $L=J\oplus K$ is the result of composing
    sequences $J$ and $K$ by order of phase number.

    \item  The occurrence of a descriptor $D$ at places where one would expect a
    sentence means that $D$ is true w.r.t. the current computation $C$.
    Sequences of descriptors are truth-evaluated conjunctively. So,
    $L$ is \textit{true/false} iff all/some of its elements are true/false.
    $J \ra K$ means that if $J$ is true then $K$ is true.
\end{enumerate}
Let us consider a computation $C$, consisting of $k$ phases,
and a milestone $\mu_j\,(j\geq 1)$.
Assume that $C$ goes for $m\geq 0$ times across $\mu_j$;
then, its \textit{history} $H_j$ is the sequence of descriptors of the form
\begin{equation}
    H_j
    =
    (p_1,j,h_1,\Delta_1),
    \ldots,
    (p_m,j,h_m,\Delta_m)
\end{equation}
where $\Delta_i$ is $+1$ if $i$ is odd and is $-1$ if $i$ is even
(since a milestone is always first crossed from the left),
and where $h_i$ is the state of $M$ when it   crossed $\mu_j$
for the $i$-th time. The sequence is empty if $m=0$.
By definition, $H_0$ begins with $(1,0,0,+1)$,
and it continues (and ends) with the descriptor $(p,0,h,-1)$
iff $M$ halts by trying to move left from cell 1.
So, if $M$ accepts after $k$
phases, we have $H_0=(1,0,0,+1),(k,0,1,-1)$.
If $C$ visits $r$ blocks, then its  \textit{history}  is the sequence
\begin{equation}
    H=H_0,H_i,\ldots,H_{r+1}
\end{equation}
where only $H_{r+1}$ is empty.

We call $H_j^+$ the sub-sequence of $\{H_j\}$ consisting of the
descriptors ending with +1, and $H_j^-$ the sub-sequence of
descriptors ending with -1. So we have $H_j=H_j^+\oplus H_j^-$.
The \textit{in-history} $\mathit{INH}_j$ of block $B_j$ is $H_j^+\oplus H_{j+1}^-$, and,
symmetrically, its \textit{out-history} $\mathit{OUTH}_j$ is $H_j^-\oplus H_{j+1}^+$.
Finally, the \textit{history  of block $B_j$} is given by
$BH_j=\mathit{INH}_j\oplus\mathit{OUTH}_j$.
A \textit{story} $S$ is a guess on a history.
Notations like $S_j,S_j^-,\mathit{OUTS}_j,\ldots$ and terms like \textit{story about milestone} $\mu_j$,
\textit{out-story} of block $B_j$, etc.
are the defined analogously to their \textit{his}torical counterparts.

\begin{figure}
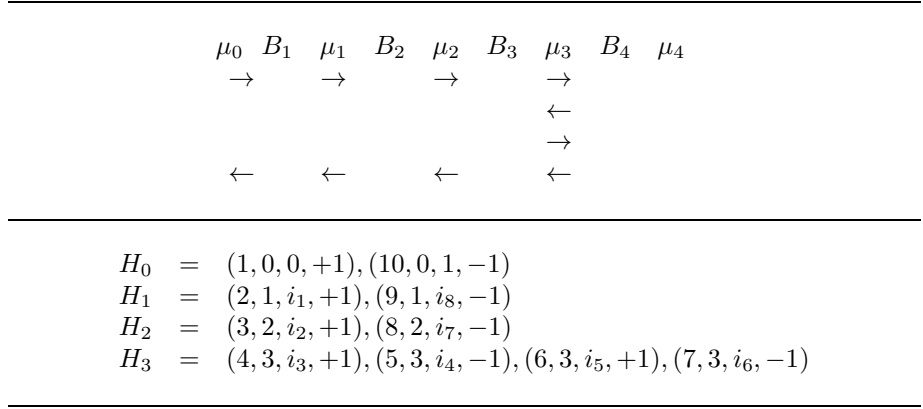

    \LINE
    \[
        \begin{array}{ccccccccc}
            \mu_0\ \ B_1&\mu_1&B_2&\mu_2&B_3&\mu_3&B_4&\mu_4&\\
            \ra\ \ \ &\ra&&\ra&&\ra\\
            &&&&&\la\\
            &&&&&\ra\\
            \la\ \ \ &\la&&\la&&\la\\
        \end{array}
    \]
    \LINE 	    	      		    	
    \[
        \begin{array}{rcl}
            H_0&=&(1,0,0,+1), (10,0,1,-1)\\
            H_1&=&(2,1,i_1,+1), (9,1,i_8,-1)\\
            H_2&=&(3,2,i_2,+1), (8,2,i_7,-1)\\
            H_3&=&(4,3,i_3,+1), (5,3,i_4,-1), (6,3,i_5,+1), (7,3,i_6,-1)\\
        \end{array}
    \]
    \LINE
    \caption{A typical history for the blocks of a \TM $M$.}\label{figure}
\end{figure}

\paragraph{Example} Assume that in a given computation $\mathit{C}$,
$M$ moves right until $B_4$, then it oscillates twice between $B_4$ and $B_3$,
and, finally, it goes left until cell 1 and accepts.
This behaviour and the related histories are sketched out in Fig.~\ref{figure}.

\section{Construction of M*}

To determine whether a given accepting story coincides with a history,
we need to introduce two \NTM.
The first one, called \texttt{phase}, takes as input an ``incoming" and
an ``outgoing" descriptor, as well as a string, and attempts to
simulate the operation of $M$ on a given block during a given phase.
The second one, \texttt{check}, works on a block by iteratively
calling \texttt{phase} and checking that a possible story of a block is
coherent across all of its phases.
Our \NTM $M^*$ works by guesswork: it makes up a story for the whole tape
(including how the tape is arranged in blocks),
and calls \texttt{check} on all of the blocks to verify whether
the story is coherent.
At the end of this section, we will show that the $M^*$ is able
to guess the history correctly.

A \NTM \texttt{phase} is employed to simulate
the behaviour of $M$ during a phase on a generic block. It is so defined:
\begin{enumerate}
    \item
    Given an input of the form
    \[
        (p,j,i,\Delta),(p+1,j^*,i^*,\Delta^*),X,
    \]
    it starts operating in the state $i$ on a string
    of the form $\langle X\rangle $, immediately
    at the right of $\langle$ for $\Delta=+1$, or at the left of $\rangle$
    for $\Delta = -1$.
    The symbols $\langle$ and $\rangle$ are not in the tape alphabet of $M$

    \item
    The machine simulates faithfully the steps of $M$,
    so each nondeterministic choice made by $M$
    causes (nondeterministically) different computations by
    \texttt{phase}.

    \item
    \texttt{phase}
    stops the simulation if $M$ halts or if, after a left/right move by
    $M$, it scans $\langle$ or $\rangle$.
    Let $\langle X^*\rangle$ be the string produced by the current computation of \texttt{phase}.
    At this point, \texttt{phase} decides whether it will accept or reject.

    \item
    \texttt{phase} rejects when one of the following conditions is verified:
    if it scans a symbol of $X^*$; if its state is not $i^*$;
    if $\langle$ is scanned, but $\Delta^*=+1$;
    and if $\rangle$ is scanned, but $\Delta*=-1$.

    \item
    In all other cases \texttt{phase} accepts, and returns the string $X^*$.
\end{enumerate}
Notice that  different  values for $X^*$ may be returned by the  computations
of \texttt{phase}.

\begin{lemma}\label{lemma_phase}
    Assume that $D$ and $D^*$ are associated with phases $p$ and $p+1$,
    and that they occur respectively in the in- and out-histories
    of $B_j$; assume further that $X$ is stored in $B_j$ at the beginning
    of  phase $p$. Then \texttt{phase} accepts and returns a content $X^*$ of $B_j$
    iff we have $D\ra D^*$.
\end{lemma}
\proof This follows immediately by construction of \texttt{phase}.

\begin{lemma}\label{lemma_check}
    Let a story be given.
    A \NTM \textbf{check} can be defined which accepts $\mathit{BS}_j$
    iff we have $\mathit{INS}_j\ra\mathit{OUTS}_j$
\end{lemma}
\proof The initial content $X(j,0)$ of $X_j$ consists of a string of $n$
zeroes if $j>2$. In $X(1,0)$ we find either the first $P$ symbols of $w$
if $P<|w|$, or $w$ followed by $P-|w|$ 0s. $X(2,0)$ begins with the part of
$w$ not stored in $B_1$ (if any), followed by a string of zeroes.

\NTM \texttt{check} works by iterating calls to \NTM \texttt{phase} :
\begin{enumerate}
    \item
    \texttt{check} calls \texttt{phase} with the following input:  the $(2p-1)$-th and $2p$-th descriptors
    of $\mathit{BS}_j$, and $X(j,p-1)$.

    \item
    If \texttt{phase} rejects, then \texttt{check} rejects too, and stops operating.

    \item
    If \texttt{phase} accepts and returns $X^*$,
    \texttt{check} puts $X(j,p)=X^*$ and starts the $(p+1)$-th repetition.
\end{enumerate}
If the last repetition of \texttt{phase} accepts, then \texttt{check}
also accepts; else, it rejects. This ends the definition of \texttt{check}.


We are now in the position to define the \mbox{\NTM $M^*$}.
Assume that $M$ accepts. Our \NTM will simulate $M$ as follows:
\begin{enumerate}
    \item{}
    $M^*$ produces a guess for the values of the time $n^2$,
    the length $P$ of the first block, the total number of visited blocks $r$,
    and the number of phases $k$.

    \item
    $M^*$ produces a guess for an accepting story
    $S=S_0,\ldots,S_{r+1}$. Since $S$ is accepting we have $S_0=((1,1,0,1),(k,1,1,-1))$
    and $S_{r+1}$ is empty.

    \item
    Next, $M^*$ calls \texttt{check} $r$ times with input
    $\mathit{BS}_j$ ($1\leq j\leq r$).

    \item
    If any call to \texttt{check} rejects, then $M^*$ rejects too; otherwise, $M^*$ accepts.
\end{enumerate}

\begin{lemma} \label{lemma_M*}
    If all calls to \texttt{check} accept,
    then $S$ is an accepting history.
\end{lemma}
\proof From lemma \ref{lemma_check} and from the hypothesis of this lemma,
the following implications are all true
\begin{equation}
    \begin{array}{l}
        S_0^+\wedge S_1^-\ra S_0^-\wedge S^+_1\\
        S_1^+\wedge S_2^-\ra S_1^-\wedge S^+_2\\
        \ldots\\
        S_j^+\wedge S_{j+1}^-\ra S_j^-\wedge S^+_{j+1}\\
        \ldots\\
        S_r^+\wedge S_{r+1}^-\ra S_r^-\wedge S^+_{r+1}\\
    \end{array}
\end{equation}
Now, note that each $S_{j}^-$\ ($1\leq j\leq r$)
occurs in the antecedent of the $j$-th implication and in the succedent of
the $(j+1)$-th
implication;
while each $S_{j}^+$
occurs in the succedent of the $j$-th,  and in the antecedent of $(j+1)$-th one. Thus all $S_j^{\pm}$ can be eliminated. Note further that $S_{r+1}^-$ and $S^+_{r+1}$
are absent (empty). Thus the above reduces to
 $S_0^+\ra S_0^-$. Since $S_0^+=(1,1,0,+1)$ is true by definition, we have that
 $S_0^-=(k,1,1,-1)$ is true. Hence, since all its descriptors are true, $S$
is an accepting history.

BV\section{Complexity}

Let us begin by analysing the space required by $M^*$.
This \NTM needs space for two activities: simulations employing \texttt{phase},
and storing the story $S$. The former works on blocks, so it clearly requires $O(n)$.
Since $S$  consists of $k$ descriptors, and the length of each of them
is $\leq c$, for a constant $c$ depending  on $M$,
we have $|S|\leq ck$.
The part of the theorem regarding space follows from the next lemma,
and the fact that $k\leq n$ implies that $|S|$ is also $O(n)$.

\begin{lemma}
    For each accepting computation $C$ there is a partition $\pi_P$
    such that its history consists of $k\leq n$ phases.
\end{lemma}

\proof
Ad absurdum. Assume that for all $P$ we had a number of phases $k(P)$ such
that $k(P)>n$. Since the overall number of moves across all boundaries is
$\leq n^2$, and since each boundary is a milestone for precisely one partition $\pi_P$,
we would have
\begin{equation}
    \sum_{P\leq n}k(P)\leq n^2.
\end{equation}
However the hypothesis ad abs. says that for each $\pi_P$ we have $k(P)>n$; that is
\begin{equation}
    \sum_{P\leq n}k(P)> n^2.
\end{equation}
This proves the lemma.

We conclude the proof of theorem \ref{theorem} by analysing the time
employed by $M^*$:
\begin{enumerate}
    \item Time for the guesses is obviously linear.

    \item By storing in the finite control of \texttt{phase}
    a description of $M$, we may arrange that it takes a time  linear
    in the time spent by $M$ (a constant number of moves for each
    simulation  of a step by $M$).
    Since each phase is simulated once, the overall time consumed by all
    calls to \texttt{phase} is $O(n^2)$.

    \item We have to add a time $O(n)$ for the $r\leq n$
    calls by $M^*$ to \texttt{check} and $O(n)$
    for the $k$ calls by \texttt{check} to \texttt{call}.
\end{enumerate}
By summing up these amounts, we obtain a time $an^2$, for some constant $a$
depending on the \NTM $M$.


\begin{thebibliography}{99}

\bibitem{HopUll68}  Hopcroft, J. E., and Ullman, J.D.: \textit{Relations between
time and tape complexities} in J. Assoc. Comp. Mach. 15(1968) 414--427.

\bibitem{HPV}  Hopcroft, J. E., Paul, W.J. and Valiant, L.G.:
\textit{On time vs. space}
  in J. Assoc. Comp. Mach. 24.2(1977) 332--337.

\bibitem{HU79}  Hopcroft, J. E., and Ullman, J.D.: \textit{Introduction to automata theory,
languages and computations.} Addison-Wesley, 1979.

\bibitem{IbarraMoran}  Ibarra, O.H. and Moran,S.:
\textit{Some time-space tradeoffs concerning single-tape and offline TMs.} SIAM J. Computing, 12.2(1983) 388--394.

\end{thebibliography}
 \end{document}